\documentclass[10pt,twocolumn,letterpaper]{article}
\usepackage[margin=0.75in,columnsep=0.25in]{geometry}
\usepackage{times,graphicx,amsmath,amssymb,booktabs,multirow,algorithm,algorithmic,xspace}
\usepackage[dvipsnames]{xcolor}
\usepackage[numbers,sort&compress]{natbib}
\usepackage[font=small,labelsep=period]{caption}
\usepackage[font=footnotesize]{subcaption}
\usepackage[shortlabels,inline]{enumitem}
\usepackage[hyphens]{url}
\usepackage[hidelinks]{hyperref}
\usepackage[capitalize]{cleveref}
\pdftrailerid{}
\hypersetup{pdfauthor={},pdfsubject={},pdfkeywords={},pdfcreator={},pdftitle={GARD: Gradient-Guided Adversarial Representation Defense against Backdoors in Multimodal Contrastive Learning Models}}
\title{GARD: Gradient-Guided Adversarial Representation Defense against Backdoors in Multimodal Contrastive Learning Models}
\author{Yuhong Wang \quad Yujun Cai \quad Lei Feng \quad
Songze Li\thanks{Corresponding author: \texttt{songzeli@seu.edu.cn}.}\\[0.4em]
{\normalsize Southeast University}\\[0.3em]
{\small \texttt{yuhongwang@seu.edu.cn}, \texttt{songzeli@seu.edu.cn}}}
\date{}
\begin{document}
\maketitle
\setlength{\textfloatsep}{6pt plus 0pt minus 0pt}
\setlength{\dbltextfloatsep}{6pt plus 0pt minus 0pt}
\makeatletter
\renewcommand{\paragraph}{\@startsection{paragraph}{4}{\z@}%
  {1.0ex plus 0.5ex minus 0.2ex}%
  {-1em}%
  {\normalfont\normalsize\bfseries}}
\makeatother
\captionsetup[table]{skip=4pt}  %
\captionsetup[figure]{skip=4pt} %
\setlength{\floatsep}{4pt plus 0pt minus 0pt}

\begin{abstract}
Multimodal contrastive learning models such as CLIP are vulnerable to backdoor
attacks with diverse trigger structures, ranging from localized patches to
globally distributed perturbations, with recent attacks increasingly optimizing
these triggers in the representation space. Existing post-hoc defenses based on clean
fine-tuning may leave backdoors active, while trigger-inversion methods depend
on accurately estimating the attack target and trigger. We observe that
backdoored encoders exhibit structure-dependent representation sensitivity:
global perturbations preferentially expose distributed backdoors, whereas
local patch perturbations are more effective for localized backdoors. Based on
this observation, we propose Gradient-Guided Adversarial Representation
Defense (GARD), a target- and trigger-agnostic post-hoc defense. GARD optimizes
complementary global and local adversarial probes to maximally displace
clean-image representations, and then realigns the perturbed representations
with detached clean anchors while preserving image–text alignment. The
resulting model requires no additional computation at inference time. Across
six visual backdoor attacks and two CLIP visual backbones, GARD reduces the
attack success rate below 1\% in all 12 evaluated attack–backbone combinations
while maintaining competitive clean accuracy. These results demonstrate the
effectiveness of adversarial representation probing for repairing backdoored
multimodal contrastive encoders.
\end{abstract}

\section{Introduction}
Multimodal contrastive learning (MCL) models, such as CLIP, have demonstrated
remarkable capabilities in learning transferable visual representations by
aligning images and texts in a shared embedding space~\cite{radford2021clip,jia2021align}.
These pretrained models have become essential components in many downstream
vision--language applications~\cite{qian2024online,mokady2021clipcap,barraco2022clipcaption,cho2022finegrained,eslami2023pubmedclip,li2022blip}, where their reliability and security are
increasingly important. However, recent studies have shown that MCL models are
vulnerable to backdoor attacks, where an attacker can implant hidden behaviors
during training such that images containing specific triggers are associated
with attacker-defined concepts while maintaining normal performance on clean
samples~\cite{jia2022badencoder,bansal2023cleanclip}. Unlike conventional
backdoors in classification models that primarily manipulate decision
boundaries, backdoors in MCL models compromise the semantic alignment between
visual and textual representations, creating new challenges for model
purification.

Existing defenses against backdoored MCL models mainly follow two paradigms:
clean-data-based purification and trigger-oriented recovery. Clean fine-tuning
methods attempt to restore benign image--text alignment using clean
data~\cite{bansal2023cleanclip}, but they provide limited supervision on
trigger-induced behaviors and may fail to remove hidden backdoor associations.
Trigger inversion methods instead exploit adversarial behaviors to estimate
attack targets or recover trigger patterns before
purification~\cite{sun2026invertune}. However, their effectiveness relies on
accurate target estimation and assumptions about the trigger structure. This
limitation becomes increasingly significant as recent attacks move beyond
handcrafted patterns toward optimized triggers with diverse spatial
structures, including localized patches and globally distributed
perturbations. Therefore, developing a defense strategy that does not rely on
specific knowledge of the attack remains an important challenge.

Recent observations on the connection between adversarial behaviors and
backdoor attacks suggest that compromised models may contain abnormal
sensitivity patterns that can be exploited without directly observing the
original trigger~\cite{sun2026invertune}. This motivates us to revisit the
objective of backdoor defense. Instead of recovering the unknown trigger
itself, an alternative direction is to identify the representation
vulnerabilities that enable different triggers to manipulate the learned
semantic space. Since backdoor attacks ultimately alter image representations
toward attacker-specified semantics, these vulnerabilities may provide a more
general defense target than the trigger patterns in the input space.

To investigate this possibility, we first study whether backdoored MCL models
exhibit structure-dependent representation vulnerabilities. Our analysis
reveals that different trigger structures produce distinct adversarial
sensitivity patterns: globally distributed backdoors are more effectively
exposed by global perturbations, whereas localized triggers are better revealed
through local perturbation probes. More importantly, such vulnerabilities can
be identified without access to the original trigger or attack target,
suggesting that adversarial perturbations can serve as a trigger-agnostic
mechanism for probing backdoor behaviors.

\begin{figure*}[t]
    \centering
    \includegraphics[width=\textwidth]{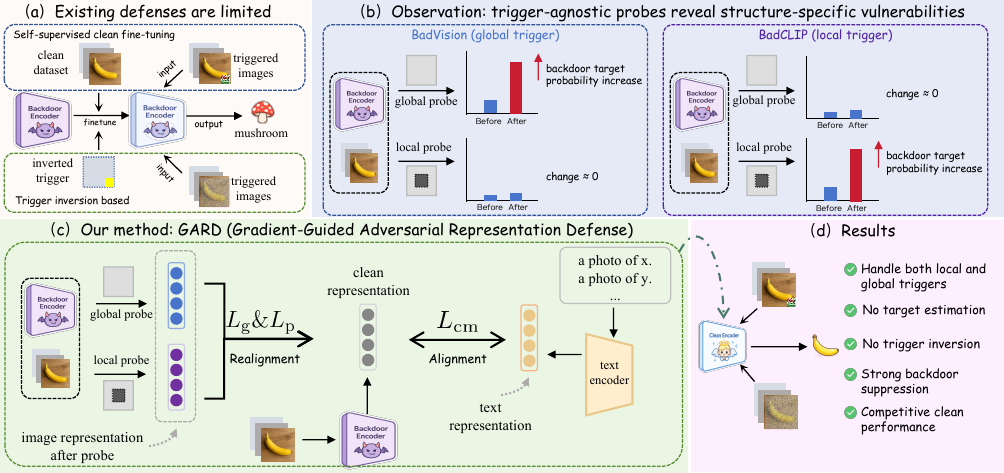}
    \caption{Overview of GARD. \textbf{(a)} Representative post-hoc defenses
    rely on clean contrastive fine-tuning, which cannot directly suppress
    backdoor behaviors under trigger-like perturbations, or use adversarial
    simulation for target estimation and trigger reconstruction, making the
    subsequent purification sensitive to trigger mismatch.
    \textbf{(b)} Trigger-agnostic adversarial probing reveals
    structure-specific vulnerabilities: global perturbations effectively expose
    globally distributed backdoors (e.g., BadVision), whereas local patch
    perturbations are more effective for localized backdoors (e.g., BadCLIP).
    \textbf{(c)} GARD jointly performs complementary global and local
    adversarial probing and realigns perturbed representations with clean
    semantic anchors while preserving clean image--text alignment.
    \textbf{(d)} A single GARD framework effectively defends against both
    localized and distributed backdoor triggers without target estimation or
    trigger reconstruction.}
    \label{fig:gard_overview}
\end{figure*}

Based on these observations, we propose \emph{Gradient-Guided Adversarial
Representation Defense} (GARD), a trigger- and target-agnostic defense for
MCL models. As illustrated in Figure~\ref{fig:gard_overview}, GARD
alternates between adversarial probing and representation realignment. It
constructs complementary global and local adversarial views to cover distributed and
spatially concentrated input variations, respectively, and realigns their
representations with a detached clean-image anchor while preserving clean
image--text correspondence. This alternating procedure requires neither the
true trigger nor its target, and introduces no additional computation at
inference time.

We evaluate GARD against six backdoor attacks spanning handcrafted patches,
blended and signal-based triggers, optimized local patches, and optimized global
perturbations, using CLIP ViT-B/32 and RN50 visual encoders
~\cite{radford2021clip,dosovitskiy2021vit,he2016resnet}. A single GARD framework
effectively handles both localized and distributed triggers, reducing ASR
below $1\%$ in all completed evaluations on both backbones while retaining
competitive clean accuracy. The ablations explain this broad coverage: local representation
alignment is critical for BadCLIP, whereas global representation alignment is
critical for BadVision, confirming that the two adversarial views address
complementary trigger structures.

Our contributions are summarized as follows:
\begin{itemize}
    \item We formulate backdoor purification as trigger-agnostic adversarial
    probing followed by representation realignment, avoiding both target-class
    estimation and explicit trigger recovery.
    \item We develop complementary global and local perturbation modules that
    probe distributed and spatially concentrated representation vulnerabilities
    under a unified training framework.
    \item We evaluate GARD across six attacks and two visual backbones. The
    results demonstrate effective backdoor removal across diverse trigger forms,
    and the ablations verify the complementary roles of two adversarial
    views.
\end{itemize}

\section{Related Work}

\paragraph{Backdoor attacks on vision--language representations.}
The shared embedding space of MCL models creates a
distinct attack surface: a small set of poisoned image--text pairs can associate
an image trigger with an attacker-chosen textual concept. Early studies showed
that such associations can survive transfer to downstream tasks and zero-shot
classifiers~\cite{jia2022badencoder,bansal2023cleanclip}. Recent attacks make
this association harder to remove by optimizing the trigger itself. BadCLIP
jointly guides the visual trigger with target-side visual and textual
embeddings, so that triggered samples approach the target semantics in both
modalities~\cite{liang2024badclip}. BadVision instead compromises a reusable
vision encoder using an imperceptible, globally distributed perturbation
optimized toward a target representation, and further trains the encoder to
distinguish the valid trigger from unrelated adversarial noise
~\cite{liu2025badvision}. These developments expand the trigger space from
fixed patches and blending patterns to optimized local or distributed
perturbations. They also motivate defenses that do not assume a particular
trigger geometry.

\paragraph{Backdoor defenses for multimodal contrastive learning models.}
Existing defenses intervene through detection, pre-training, post-hoc
fine-tuning, or inference. DECREE detects compromised encoders without
repairing them~\cite{feng2023decree}. SAFECLIP separates likely clean and risky image--text pairs during
pre-training and applies cross-modal learning only to the progressively
expanded safe subset~\cite{yang2024safeclip}; this setting presumes control over
the original training process. Post-hoc purification instead uses clean-data
fine-tuning and adversarial augmentation~\cite{bansal2023cleanclip,kuang2024abd}.
The connection between adversarial perturbations
and backdoor behavior is already well recognized in backdoor defense.
InverTune exploits this connection by generating a universal adversarial
perturbation to elicit target-associated responses from the compromised model;
it then estimates the attack target, reconstructs a latent trigger, and applies
clustering-guided purification~\cite{sun2026invertune}. Its purification signal
nevertheless depends on target estimation and trigger inversion.
Representation-decomposition analyses further show that local
patch and global perturbation attacks affect different CLIP components, and
support component-level repair or filtering at inference
~\cite{he2025closer}. In contrast, GARD is a post-hoc defense that requires
neither target estimation nor trigger recovery. Rather than treating the
adversarial--backdoor connection itself as a contribution, GARD uses
target-free adversarial optimization to probe the compromised encoder under
complementary global and local perturbation spaces, and directly realigns both
adversarial representations to clean semantic anchors.
\section{Preliminary Study}
\label{sec:preliminary}

We first examine whether trigger-agnostic adversarial probing exposes
structure-specific backdoor vulnerabilities. We use the same 256 non-target
ImageNet images to probe two backdoored ViT-B/32 encoders: BadVision, whose
trigger is distributed over the image, and BadCLIP, whose optimized
$16\!\times\!16$ trigger is placed at the bottom-right corner. For each clean
image, we independently optimize (i) an $\ell_\infty$-bounded global
perturbation and (ii) a $32\!\times\!32$ replacement patch at the image center.
Both perturbations maximize the representation deviation from the clean image
without using the attack target or the real trigger. Because the two
perturbation spaces have different capacities, we measure the increase in
target-class probability relative to a matched random initialization of each
perturbation.

\begin{figure}[t]
    \centering
    \includegraphics[width=\columnwidth]{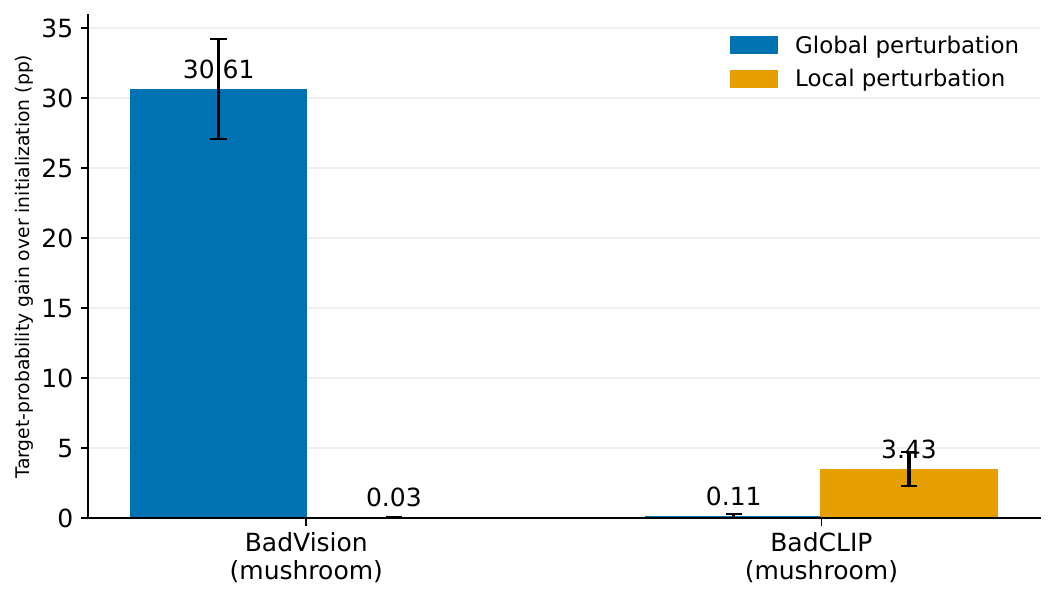}
    \caption{Trigger-agnostic adversarial sensitivity under different spatial
    supports. Bars report the increase in target-class probability produced by
    optimization over a matched random initialization; error bars denote 95\%
    bootstrap confidence intervals over images. The mean target-class
    probabilities at initialization are 0.016\% (global) and 0.027\% (local)
    for BadVision, and 0.0002\% (global) and 0.0036\% (local) for BadCLIP.}
    \label{fig:preliminary_locality}
\end{figure}

\begin{figure*}[t]
    \centering
    \includegraphics[width=0.90\textwidth]{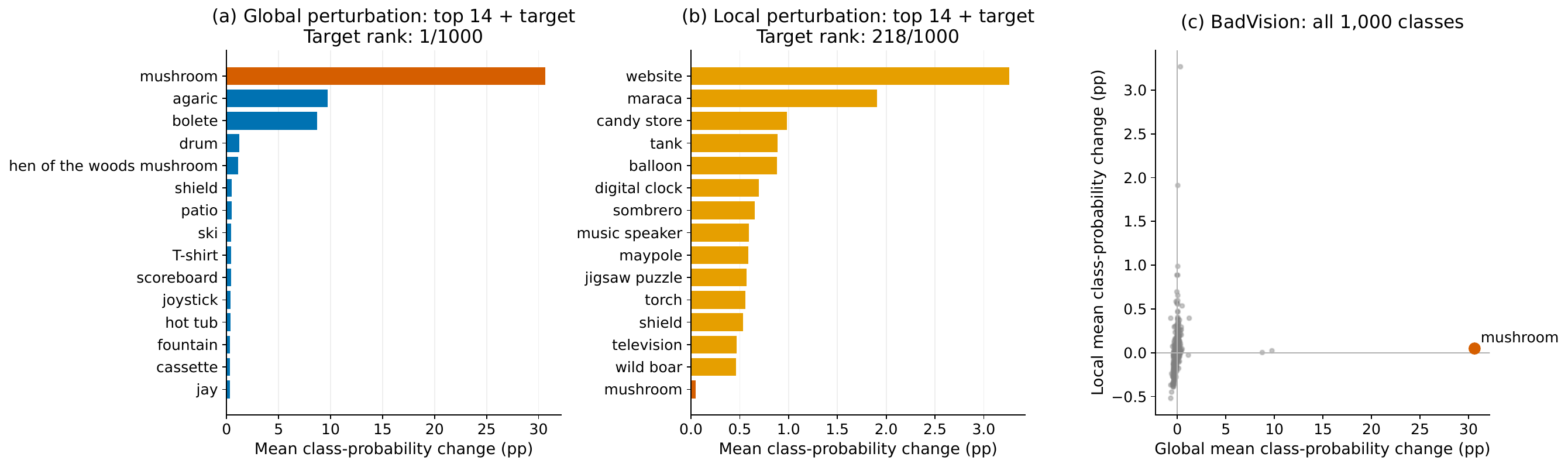}
    \vspace{2mm}

    \includegraphics[width=0.90\textwidth]{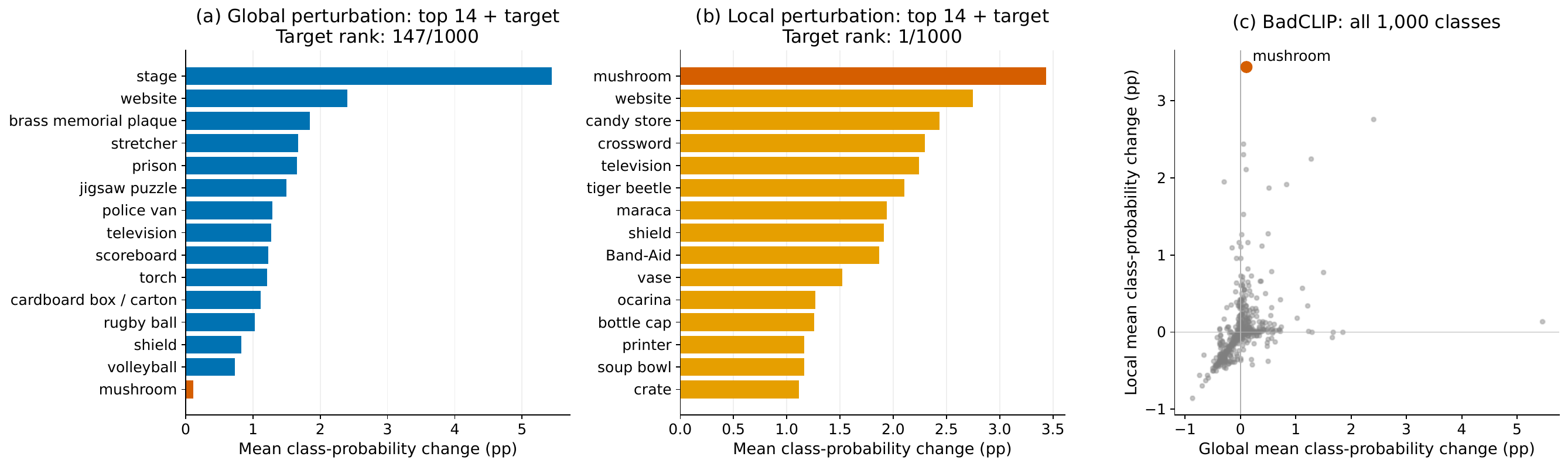}
    \caption{Class-wise responses to trigger-agnostic global and local
    perturbations on BadVision (top) and BadCLIP (bottom), with both attacks
    targeting \emph{mushroom}. The bar plots show the classes with the largest
    mean probability increases relative to clean inputs, together with the
    implanted target; the scatter plots show all 1,000 ImageNet classes. For
    BadVision, the target ranks first under the global perturbation but 218th
    under the local perturbation. For BadCLIP, the target ranks first under the
    local perturbation but 147th under the global perturbation.}
    \label{fig:preliminary_classwise}
\end{figure*}

\paragraph{Observation 1: Perturbation-specific backdoor sensitivity.}
Figure~\ref{fig:preliminary_locality} shows that different types of backdoors
exhibit distinct sensitivities to different forms of adversarial
perturbations. For BadVision, the
global perturbation increases the target probability by 30.61 percentage
points over its initialization, compared with 0.03 points for the local
perturbation. The trend reverses for BadCLIP: the local perturbation produces a
3.43-point increase, compared with 0.11 points for the global perturbation.
The class-wise comparison in Figure~\ref{fig:preliminary_classwise} further
shows that the corresponding target response is selective relative to the
other 999 classes. These results suggest that backdoors leave a structure-dependent vulnerability
footprint in the representation space: distributed and patch-based triggers
are preferentially exposed by global and local perturbations, respectively.
This observation is also consistent with the failure of a localized
target-identification probe on globally distributed triggers. Controlled
experiments in the appendix show that this structural mismatch
causes InverTune to identify an incorrect target, which in turn severely
compromises subsequent backdoor removal.

\paragraph{Observation 2: Exact trigger geometry is unnecessary.}
The experiment further shows that a surrogate perturbation can expose the
model's target-associated backdoor vulnerability even when its location and
size differ from those of the genuine trigger. The genuine BadCLIP trigger is a
$16\!\times\!16$ patch at the bottom-right corner, whereas our surrogate is a
$32\!\times\!32$ patch optimized at the image center. Despite this mismatch in
both location and size, Figure~\ref{fig:preliminary_classwise} shows that the
surrogate makes the implanted target the class with the largest mean
probability increase among all 1,000 ImageNet classes.
Exposing a localized backdoor vulnerability therefore does not require exact
knowledge of the trigger pattern, location, or scale.

Our preliminary studies reveal that different trigger structures induce distinct
sensitivity patterns in the model's representation space, which can be identified
through trigger-agnostic adversarial probing. Motivated by these findings, we propose
GARD to use complementary global and local adversarial views that cover different spatial
supports, without estimating the attack target or reproducing the implanted
trigger.

\section{Method}

\subsection{Problem Formulation}

We consider an MCL model consisting of an image encoder $f_{\theta}$ and a
text encoder $g_{\phi}$, which map images and texts into a shared
$d$-dimensional representation space. Given an image $x$ and text $t$, their
normalized representations are
\begin{equation}
    z_x =
    \frac{f_{\theta}(x)}
    {\lVert f_{\theta}(x) \rVert_2},
    \qquad
    z_t =
    \frac{g_{\phi}(t)}
    {\lVert g_{\phi}(t) \rVert_2}.
\end{equation}
Cross-modal matching is performed using their cosine similarity.

We study a post-hoc defense setting in which the defender has access only to
the backdoored model and a small clean image--text dataset
$\mathcal{D}_c=\{(x_i,t_i)\}_{i=1}^{N}$. The poisoned training data, trigger,
and target class are unavailable. Our goal is to neutralize the implanted
backdoor while preserving the model's utility on clean data.

\subsection{Overview of GARD}

As illustrated in Figure~\ref{fig:gard_overview}, GARD mitigates backdoor
attacks in MCL models through a two-stage process: complementary global and
local adversarial probing, followed by clean-anchored representation
realignment. Since backdoor attacks are designed to preserve clean utility, an
untriggered clean image generally remains aligned with its paired text and thus
provides a reliable reference for its intended semantics. GARD treats this
clean representation as a fixed anchor and realigns the deliberately displaced
representations toward it, thereby suppressing the vulnerable directions
through which trigger-like perturbations redirect an image toward
attacker-specified semantics.

\subsection{Gradient-Guided Global Perturbation}

Backdoor triggers are not always confined to localized patterns but may also
manifest as distributed perturbations over the entire image. To expose such
global vulnerabilities, we construct a bounded perturbation that maximally
displaces the image representation from its clean counterpart.

Given a clean image $x\in[0,1]^{H\times W\times3}$, we generate a globally
perturbed view
\begin{equation}
x^{g}=\operatorname{clip}_{[0,1]}(x+\delta),
\end{equation}
where $\|\delta\|_{\infty}\leq\epsilon$. The perturbation is obtained by
maximizing the representation discrepancy between the clean and perturbed
images:
\begin{equation}
\delta^{\star}
=
\operatorname*{argmax}_{\|\delta\|_{\infty}\leq\epsilon}
\left[
1-
\operatorname{sim}
\bigl(
\bar f_{\theta}(x+\delta),
\bar f_{\theta}(x)
\bigr)
\right],
\label{eq:global_perturbation}
\end{equation}
where $\bar f_{\theta}$ denotes the $\ell_2$-normalized image representation
and $\operatorname{sim}(\cdot,\cdot)$ is cosine similarity. We approximately
solve Eq.~\eqref{eq:global_perturbation} using projected sign-gradient updates
under the $\ell_{\infty}$ constraint.

Unlike trigger inversion, the optimized perturbation is not intended to recover
the unknown trigger pattern. Instead, it searches for perturbation directions
that induce the largest representation deviation within a bounded perturbation
budget. Because backdoored models exhibit abnormal sensitivity to
trigger-related perturbations, these directions preferentially reveal the
backdoor-induced vulnerable representations, providing effective supervision
for the subsequent representation realignment.

\subsection{Gradient-Guided Local Patch Perturbation}

While the global perturbation effectively explores distributed
representation vulnerabilities, many practical backdoor attacks rely on
spatially localized triggers. To expose such localized vulnerabilities, we
construct a second adversarial view by replacing a square image region with a
learnable patch.

Given a clean image $x$, let $\mathcal{P}(x,p,\ell)$ denote the operation of
pasting a learnable patch $p$ at location $\ell$. The patch is optimized to
maximize the representation discrepancy between the patched and clean images:
\begin{equation}
\begin{aligned}
p^{\star}
&=
\operatorname*{argmax}_{p\in[0,1]^{s\times s\times3}}
\\
&\quad
\left[
1-
\operatorname{sim}
\bigl(
\bar f_{\theta}(\mathcal{P}(x,p,\ell)),
\bar f_{\theta}(x)
\bigr)
\right],
\end{aligned}
\label{eq:patch_perturbation}
\end{equation}
where $s$ denotes the patch size. We approximately solve
Eq.~\eqref{eq:patch_perturbation} using projected sign-gradient updates while
keeping the patch values within the valid image range.

Unlike trigger inversion methods that attempt to reconstruct the attack
pattern, the optimized patch is only required to maximally disrupt the image
representation. By concentrating perturbations within a local region, the
patch effectively probes spatially localized vulnerabilities that may be
overlooked by globally constrained perturbations. Together with the global
perturbation, it provides a complementary view for exposing abnormal
representation sensitivity in backdoored models.

\subsection{Representation Realignment}

Given the generated global and local adversarial views, GARD fine-tunes the
backdoored model to achieve two objectives: preserving the original
cross-modal alignment and restoring perturbed image representations to their
clean semantics.

To preserve the model's visual--language capability, we employ the standard
bidirectional contrastive loss on clean image--text pairs
~\cite{radford2021clip}. Given a mini-batch of $B$ samples, the image-to-text
loss is defined as
\begin{equation}
    \mathcal{L}_{\mathrm{i2t}}
    =
    -\frac{1}{B}
    \sum_{i=1}^{B}
    \log
    \frac{
        \exp\left(\tau z_{x_i}^{\top}z_{t_i}\right)
    }{
        \sum_{j=1}^{B}
        \exp\left(\tau z_{x_i}^{\top}z_{t_j}\right)
    }.
    \label{eq:i2t}
\end{equation}
Similarly, the text-to-image loss is
\begin{equation}
    \mathcal{L}_{\mathrm{t2i}}
    =
    -\frac{1}{B}
    \sum_{i=1}^{B}
    \log
    \frac{
        \exp\left(\tau z_{t_i}^{\top}z_{x_i}\right)
    }{
        \sum_{j=1}^{B}
        \exp\left(\tau z_{t_i}^{\top}z_{x_j}\right)
    }.
    \label{eq:t2i}
\end{equation}
The clean cross-modal alignment objective is then
\begin{equation}
    \mathcal{L}_{\mathrm{cm}}
    =
    \frac{1}{2}
    \left(
        \mathcal{L}_{\mathrm{i2t}}
        +
        \mathcal{L}_{\mathrm{t2i}}
    \right),
    \label{eq:cross_modal}
\end{equation}
where $\tau$ is the learnable logit scale.

To suppress abnormal representation sensitivity revealed by adversarial
perturbations, we further align the representation of each perturbed image
with its clean counterpart. For each perturbation type
$q\in\{g,p\}$, the representation alignment loss is
\begin{equation}
    \mathcal{L}_{q}
    =
    \frac{1}{B}
    \sum_{i=1}^{B}
    \left[
    1-
    \operatorname{sim}
    \left(
    \bar f_{\theta}(x_i^{q}),
    \operatorname{sg}\!\left[\bar f_{\theta}(x_i)\right]
    \right)
    \right],
    \label{eq:consistency}
\end{equation}
where $\operatorname{sg}[\cdot]$ denotes the stop-gradient operation, which
treats the clean representation as a fixed semantic anchor during
optimization.

The overall training objective is
\begin{equation}
    \mathcal{L}_{\mathrm{GARD}}
    =
    \lambda_{\mathrm{cm}}\mathcal{L}_{\mathrm{cm}}
    +
    \lambda_{g}\mathcal{L}_{g}
    +
    \lambda_{p}\mathcal{L}_{p},
    \label{eq:gard_objective}
\end{equation}
where $\lambda_{\mathrm{cm}}$, $\lambda_g$, and $\lambda_p$ balance the three
objectives. Optimizing Eq.~\eqref{eq:gard_objective} encourages the model to
maintain clean image--text alignment while progressively restoring the
representations of both globally and locally perturbed views. Consequently,
the abnormal representation sensitivity exposed during perturbation generation
is gradually suppressed without requiring trigger recovery or target-class
estimation.

GARD is trained by alternating between perturbation generation and representation realignment.
After training, the purified MCL model is directly used for inference without 
any additional test-time computation.

\section{Experiments}
\subsection{Experimental Setup}
\paragraph{Backdoor Attacks.}
We consider BadNets~\cite{gu2019badnets}, Blended~\cite{chen2017targeted},
SIG~\cite{barni2019sig}, BadEncoder~\cite{jia2022badencoder},
BadCLIP~\cite{liang2024badclip}, and BadVision~\cite{liu2025badvision} on
OpenAI CLIP ViT-B/32 and RN50 encoders~\cite{radford2021clip}. Together, these
attacks cover handcrafted patches, blended and signal-based triggers and 
optimized triggers. Complete attack configurations are provided in the appendix.

\paragraph{Baselines.}
We compare GARD with FT, CleanCLIP~\cite{bansal2023cleanclip},
PAR~\cite{singh2026par}, and InverTune~\cite{sun2026invertune}, and also report
each backdoored model without defense.
Each defense is applied independently to the same backdoored checkpoint for
each attack--backbone pair. Following the experimental setup of InverTune, we
use 500K CC3M image--text pairs~\cite{sharma2018conceptual} for FT, CleanCLIP,
and PAR, 50K ImageNet samples for InverTune, and 20K ImageNet samples for
GARD.
FT, CleanCLIP, PAR, and GARD are trained for 10 epochs, whereas InverTune
follows its original multi-stage schedule. Additional implementation details
and defense runtimes are provided in Appendix Sections~E and I,
respectively.

\paragraph{Evaluation Metrics.}
We report clean accuracy (CA) and attack success rate (ASR) on ImageNet-1K
~\cite{deng2009imagenet}. CA is the top-1 zero-shot accuracy on clean images,
whereas ASR is the percentage of triggered non-target images classified as the
attacker-specified target. Images whose ground-truth class is already the
attack target are excluded.

\subsection{Main Results}

\begin{table*}[t]
\centering
\caption{Main results across six attacks and two visual backbones. CA (\%)
measures clean utility, and ASR (\%) measures backdoor effectiveness after
defense. The best and second-best results are shown in bold and underlined, respectively.}
\label{tab:main_results}
\small
\setlength{\tabcolsep}{5.3pt}
\begin{tabular}{ll cc cc cc cc cc cc}
\toprule
& & \multicolumn{2}{c}{BadNets}
& \multicolumn{2}{c}{Blended}
& \multicolumn{2}{c}{SIG}
& \multicolumn{2}{c}{BadEncoder}
& \multicolumn{2}{c}{BadCLIP}
& \multicolumn{2}{c}{BadVision} \\
\cmidrule(lr){3-4}\cmidrule(lr){5-6}\cmidrule(lr){7-8}
\cmidrule(lr){9-10}\cmidrule(lr){11-12}\cmidrule(lr){13-14}
Backbone & Method
& CA $\uparrow$ & ASR $\downarrow$
& CA $\uparrow$ & ASR $\downarrow$
& CA $\uparrow$ & ASR $\downarrow$
& CA $\uparrow$ & ASR $\downarrow$
& CA $\uparrow$ & ASR $\downarrow$
& CA $\uparrow$ & ASR $\downarrow$ \\
\midrule
\multirow{6}{*}{ViT-B/32}
& No Defense
& 60.03 & 99.92 & 60.18 & 99.86 & 60.08 & 99.55
& 62.94 & 99.38 & 60.58 & 99.94 & 61.30 & 99.56 \\
& FT
& 56.42 & 44.09 & 56.73 & 75.71 & 56.45 & 22.81
& 56.40 & 0.12 & 56.99 & 89.61 & 56.43 & 38.14 \\
& CleanCLIP
& 54.91 & 31.46 & 54.94 & 0.12 & 55.24 & \underline{7.96}
& 55.32 & \textbf{0.04} & 54.78 & 77.48 & 54.96 & 18.35 \\
& PAR
& 54.66 & 41.16 & 51.59 & \underline{0.08} & 54.77 & 43.16
& 51.63 & 0.07 & 53.70 & 19.59 & 51.94 & \underline{11.81} \\
& InverTune
& \underline{57.29} & \underline{6.10}
& \underline{58.33} & 1.98
& \textbf{58.57} & 9.91
& \underline{58.61} & 0.49
& \underline{57.98} & \underline{4.70}
& \underline{57.15} & 55.70 \\
& \textbf{GARD}
& \textbf{59.36} & \textbf{0.10}
& \textbf{59.23} & \textbf{0.01}
& \underline{57.61} & \textbf{0.88}
& \textbf{61.66} & \underline{0.05}
& \textbf{61.01} & \textbf{0.82}
& \textbf{60.39} & \textbf{0.63} \\
\midrule
\multirow{6}{*}{RN50}
& No Defense
& 58.94 & 99.47 & 58.99 & 99.20 & 59.11 & 98.22
& 57.23 & 98.01 & 58.35 & 99.38 & 51.16 & 84.71 \\
& FT
& \textbf{54.78} & 35.55
& \textbf{54.56} & 0.19
& \textbf{54.73} & 5.67
& \underline{54.81} & 0.13
& \textbf{56.99} & 89.68
& \textbf{54.72} & 11.49 \\
& CleanCLIP
& 52.74 & 21.95 & 52.61 & 0.10 & 52.36 & 1.18
& 52.34 & 0.08 & 52.38 & 19.97 & 52.81 & 28.74 \\
& PAR
& 49.03 & \underline{0.97}
& 49.13 & 0.16
& 48.99 & \textbf{0.17}
& 50.28 & \underline{0.02}
& 50.46 & \underline{2.50}
& 49.71 & \underline{2.10} \\
& InverTune
& 51.27 & 5.61 & 51.12 & \underline{0.08} & 50.82 & 4.86
& 51.01 & 1.04 & 54.71 & 6.06 & 51.83 & 33.45 \\
& \textbf{GARD}
& \underline{53.35} & \textbf{0.11}
& \underline{52.79} & \textbf{0.03}
& \underline{53.44} & \underline{0.45}
& \textbf{55.76} & \textbf{0.01}
& \underline{54.88} & \textbf{0.03}
& \underline{53.97} & \textbf{0.05} \\
\bottomrule
\end{tabular}

\end{table*}

As shown in Table~\ref{tab:main_results}, across all six attacks and both
visual backbones, GARD is the only defense
that reduces ASR below 1\% in every evaluated setting, demonstrating
consistent protection against different backdoor attacks. By contrast, every
baseline exhibits at least one attack on which it fails to provide reliable
protection, highlighting the limited generalization of existing defenses
across distinct trigger structures. These results validate the key intuition
behind GARD that complementary global and local adversarial probing provides
a unified defense against diverse backdoor behaviors.

Across both visual backbones, GARD maintains competitive clean accuracy while
effectively removing backdoors, demonstrating a favorable balance between
model utility and attack mitigation.

\subsection{Comparison with Matched Defense Data}
\label{sec:matched_data}
To assess whether GARD's performance stems from differences in defense data,
we rerun FT, CleanCLIP, PAR, and InverTune on ViT-B/32 using the same
class-balanced 20K ImageNet subset as GARD. We compare these results with
GARD's main-experiment results, keeping the defense-data source, sample
count, and class composition consistent across methods.

\begin{table}[t]
\centering
\caption{ASR (\%, $\downarrow$) on ViT-B/32 with matched defense data:
all defenses use the same class-balanced 20K ImageNet subset.}
\label{tab:matched_data}
\footnotesize
\setlength{\tabcolsep}{0pt}
\begin{tabular*}{\linewidth}{@{\extracolsep{\fill}}lcccccc@{}}
\toprule
Defense & BadNets & Blended & SIG & BadEncoder & BadCLIP & BadVision \\
\midrule
No Defense & 99.92 & 99.86 & 99.55 & 99.38 & 99.94 & 99.56 \\
FT & 99.73 & 90.11 & 93.22 & 5.01 & 99.91 & 82.89 \\
CleanCLIP & 99.79 & 33.40 & 90.95 & 1.01 & 99.92 & 80.79 \\
PAR & 97.95 & 0.11 & 80.07 & 25.71 & 99.86 & 77.71 \\
InverTune & 7.91 & 2.59 & 9.63 & 7.12 & 4.19 & 52.16 \\
GARD & \textbf{0.10} & \textbf{0.01} & \textbf{0.88} & \textbf{0.05} & \textbf{0.82} & \textbf{0.63} \\
\bottomrule
\end{tabular*}
\end{table}

Table~\ref{tab:matched_data} shows that GARD retains its defense advantage
when all methods use the same clean dataset. Compared with their
results using 500K CC3M pairs, FT, CleanCLIP, and PAR are substantially less
effective against several attacks in the 20K ImageNet setting, with ASR
remaining above 77\% on BadNets, BadCLIP, and BadVision. These results
suggest that their effectiveness depends on access to larger clean-data
collections, while highlighting GARD's ability to remove diverse backdoors
with a limited clean-data budget.

\subsection{Evaluation against Adaptive Attacks}
\label{sec:adaptive_attacks}
To evaluate GARD against attacks tailored to its defense mechanism, we
design two adaptive attack strategies based on BadCLIP and BadVision.
We assume that the attacker knows GARD's perturbation generation procedures,
training objectives, and hyperparameters, and uses this knowledge to optimize
backdoors that resist purification. Both strategies add a defense-resistance
loss to the base attack objective.

The first strategy, \textbf{Adap-1 (simulated purification)}, explicitly
optimizes backdoor persistence after defense. During attack training, the
attacker applies two GARD updates to a copy of the current model every five
steps and minimizes the target-class loss on triggered images using the
purified model, aiming to preserve backdoor effectiveness after defense.

The second strategy, \textbf{Adap-2 (defense-signal evasion)}, targets the
representation discrepancies that guide GARD's purification. The attacker
reuses GARD's global and local adversarial probing procedures and minimizes
the resulting discrepancies between clean and perturbed representations.
This additional loss aims to weaken the signal used by GARD for purification.
We apply both strategies to BadCLIP and BadVision
on ViT-B/32 and evaluate the resulting models with the full GARD defense.
Implementation details are provided in the appendix.

\begin{table}[t]
\centering
\caption{Adaptive attacks against GARD on ViT-B/32.
Before and After refer to evaluation before and after the full GARD defense.}
\label{tab:adaptive_attacks}
\footnotesize
\setlength{\tabcolsep}{3pt}
\begin{tabular}{lrrrr}
\toprule
& \multicolumn{2}{c}{BadCLIP} & \multicolumn{2}{c}{BadVision} \\
\cmidrule(lr){2-3}\cmidrule(lr){4-5}
Stage & CA $\uparrow$ & ASR $\downarrow$ & CA $\uparrow$ & ASR $\downarrow$ \\
\midrule
\multicolumn{5}{l}{\textit{Adap-1: Simulated purification}} \\
Before & 63.97 & 99.58 & 62.82 & 99.84 \\
After & 62.45 & 0.01 & 60.93 & 0.76 \\
\midrule
\multicolumn{5}{l}{\textit{Adap-2: Defense-signal evasion}} \\
Before & 64.65 & 97.01 & 63.98 & 99.77 \\
After & 62.59 & 0.09 & 62.28 & 0.06 \\
\bottomrule
\end{tabular}
\end{table}

Table~\ref{tab:adaptive_attacks} shows that both strategies achieve high
attack success before defense, with ASR ranging from 97.01\% to 99.84\%.
Despite explicitly targeting GARD's purification procedure and alignment
signals, these attacks are effectively mitigated. GARD reduces ASR below
1\% in all four settings while largely preserving clean accuracy. These
results demonstrate GARD's robustness to both adaptive
strategies. We believe this robustness stems from GARD's ability to continually probe
and suppress backdoor behavior during purification, rather than relying on
a fixed trigger pattern. Although individual probing steps do not aim to
recover the exact trigger, repeated probing and representation realignment
can cumulatively suppress the backdoor over successive iterations.

\subsection{Ablation Studies}

\begin{table}[t]
\centering
\caption{Ablation of the training objectives in GARD on ViT-B/32. A bullet
indicates that the corresponding objective is enabled. The best and
second-best results among the defense variants are shown in bold and
underlined, respectively.}
\label{tab:ablation}
\footnotesize
\setlength{\tabcolsep}{2pt}
\begin{tabular}{lccc cc cc}
\toprule
& \multicolumn{3}{c}{Training Objective}
& \multicolumn{2}{c}{BadCLIP}
& \multicolumn{2}{c}{BadVision} \\
\cmidrule(lr){2-4}\cmidrule(lr){5-6}\cmidrule(lr){7-8}
Variant
& $\mathcal{L}_{\mathrm{cm}}$
& $\mathcal{L}_{g}$
& $\mathcal{L}_{p}$
& CA $\uparrow$ & ASR $\downarrow$
& CA $\uparrow$ & ASR $\downarrow$ \\
\midrule
No Defense
& -- & -- & --
& 60.58 & 99.94
& 61.30 & 99.56 \\
Base
& $\bullet$ & -- & --
& \textbf{62.66} & 99.90
& \textbf{63.39} & 73.87 \\
w/o Global
& $\bullet$ & -- & $\bullet$
& \underline{62.23} & \textbf{0.31}
& \underline{62.79} & 36.63 \\
w/o Local
& $\bullet$ & $\bullet$ & --
& 61.95 & 99.84
& 60.67 & \underline{0.97} \\
Full GARD
& $\bullet$ & $\bullet$ & $\bullet$
& 61.01 & \underline{0.82}
& 60.39 & \textbf{0.63} \\
\bottomrule
\end{tabular}

\end{table}

Table~\ref{tab:ablation} validates the contribution of each training objective in GARD under the BadCLIP and BadVision attacks, which represent optimized local patch triggers and globally optimized perturbations, respectively. The Base variant, which employs only the clean cross-modal alignment objective, provides little protection, retaining ASRs of 99.90\% and 73.87\% on BadCLIP and BadVision, respectively. This indicates that preserving clean cross-modal alignment alone is insufficient to eliminate backdoor-induced representation deviations.

Removing the global representation alignment objective preserves a low ASR on BadCLIP (0.31\%) but substantially weakens the defense against BadVision, increasing the ASR to 36.63\%. This result suggests that global representation alignment primarily mitigates vulnerabilities introduced by globally optimized perturbations. In contrast, removing the local representation alignment objective leaves the defense against BadVision largely intact (0.97\%) but causes the ASR on BadCLIP to increase to 99.84\%, indicating that local representation alignment is essential for suppressing patch-based triggers.

These observations demonstrate that the two representation alignment objectives address complementary vulnerabilities induced by different trigger patterns. Optimizing both objectives jointly enables Full GARD to reduce the ASR below 1\% on both attacks simultaneously. Although the complete objective incurs a modest reduction in clean accuracy compared with the Base variant, it maintains clean performance close to that of the undefended model while providing consistently strong protection across diverse trigger types.

\subsection{Further Analysis}

\paragraph{Representation Realignment Visualization.}
\begin{figure}[t]
    \centering
    \includegraphics[width=\columnwidth]{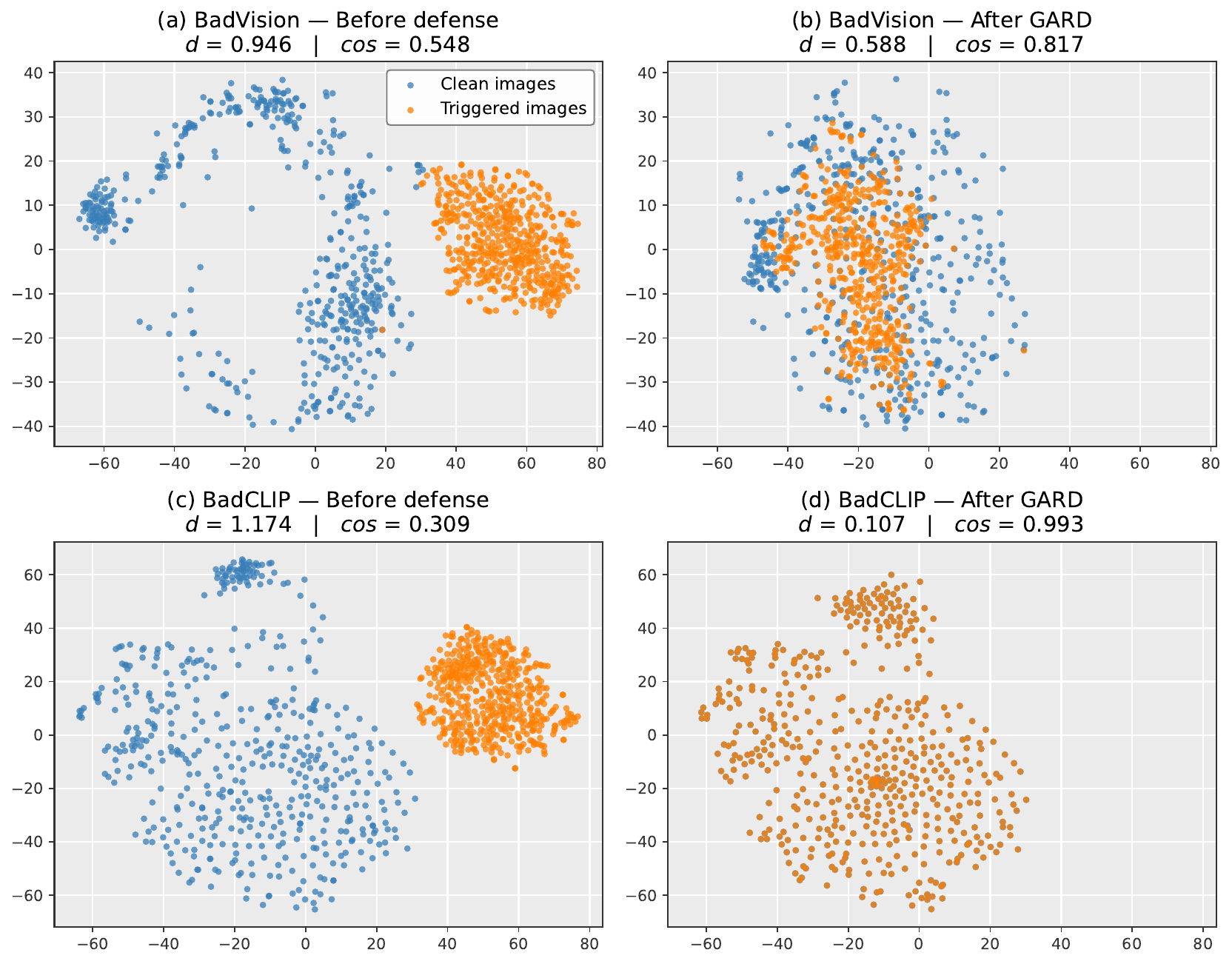}
    \caption{Paired clean and triggered ViT-B/32 representations before and
    after GARD.
    Mean Euclidean distance $d$ and cosine similarity $\mathit{cos}$ are
    computed in the original normalized feature space.}
    \label{fig:representation_tsne}
\end{figure}

Figure~\ref{fig:representation_tsne} examines whether the reduction in ASR is
accompanied by a correction of trigger-induced representation deviations.
Before defense, clean and triggered representations form clearly separated
regions for both attacks. After GARD, their distributions overlap more
substantially: for BadVision, the mean distance decreases from 0.946 to 0.588
and the mean cosine similarity increases from 0.548 to 0.817; for BadCLIP,
the corresponding values change from 1.174 to 0.107 and from 0.309 to 0.993.
These results provide representation-level evidence that GARD suppresses the
deviations induced by both globally distributed and localized triggers,
although the remaining deviation is larger for BadVision. Complete
visualizations covering all six attacks, together with clean-model
references, are provided in the appendix.

\paragraph{Sensitivity to Perturbation Parameters.}
We study the sensitivity of GARD to the global perturbation budget $\epsilon$
and the local patch size $s$ on ViT-B/32. We vary $\epsilon$ under BadVision
and $s$ under BadCLIP while keeping all other training settings fixed.

\begin{table}[t]
\centering
\caption{Sensitivity of GARD to the global perturbation budget $\epsilon$ and
local patch size $s$ on ViT-B/32. $\dagger$ denotes the default setting used
in the main experiments.}
\label{tab:parameter_analysis}
\footnotesize
\begin{tabular}{lcc}
\toprule
Parameter & CA $\uparrow$ & ASR $\downarrow$ \\
\midrule
\multicolumn{3}{c}{Global Perturbation on BadVision} \\
\midrule
$\epsilon=2/255$ & 61.77 & 0.65 \\
$\epsilon=4/255$ & 61.11 & 0.42 \\
$\epsilon=8/255^{\dagger}$ & 60.39 & 0.63 \\
$\epsilon=16/255$ & 60.04 & 0.05 \\
\midrule
\multicolumn{3}{c}{Local Patch on BadCLIP} \\
\midrule
$s=16$ & 61.27 & 6.64 \\
$s=24$ & 60.07 & 0.53 \\
$s=32^{\dagger}$ & 61.01 & 0.82 \\
$s=48$ & 60.15 & 0.06 \\
\bottomrule
\end{tabular}

\end{table}

As shown in Table~\ref{tab:parameter_analysis}, GARD maintains low ASR on
BadVision across the evaluated perturbation budgets. Increasing the budget to
$16/255$ further reduces the ASR to 0.05\%, accompanied by a decrease in CA,
reflecting a trade-off between defense strength and clean utility. For the
local perturbation, a small patch of $s=16$ is less effective, whereas all
settings with $s\geq24$ reduce the BadCLIP ASR below 1\%. This suggests that
adequate spatial support is important for exposing localized representation
vulnerabilities, while the defense does not require an exact estimate of the
trigger size. Overall, GARD remains effective over a moderate range of both
parameters, with the default settings providing consistently low ASR while
preserving clean accuracy.

\begin{table}[!ht]
\centering
\caption{Effect of clean-data size on GARD with ViT-B/32. $\dagger$ denotes
the default setting used in the main experiments.}
\label{tab:data_size}
\footnotesize
\begin{tabular}{lcccc}
\toprule
& \multicolumn{2}{c}{BadCLIP}
& \multicolumn{2}{c}{BadVision} \\
\cmidrule(lr){2-3}\cmidrule(lr){4-5}
Clean Data
& CA $\uparrow$ & ASR $\downarrow$
& CA $\uparrow$ & ASR $\downarrow$ \\
\midrule
1K
& 57.09 & 14.62
& \underline{59.81} & 87.59 \\
5K
& \underline{58.59} & 3.75
& 59.28 & 61.10 \\
10K
& 58.13 & \underline{2.10}
& 59.68 & \underline{15.91} \\
20K$^{\dagger}$
& \textbf{61.01} & \textbf{0.82}
& \textbf{60.39} & \textbf{0.63} \\
\bottomrule
\end{tabular}

\end{table}

\paragraph{Effect of Clean Data Size.}
We further evaluate GARD with ImageNet subsets containing 1K, 5K, 10K, and
20K clean images. All subsets are class-balanced, with 1, 5, 10, and 20
images per class, respectively.

Table~\ref{tab:data_size} shows that backdoor removal improves consistently as
more clean data become available. Increasing the subset from 1K to 20K images
lowers ASR from 14.62\% to 0.82\% on BadCLIP and from 87.59\% to 0.63\% on
BadVision, while the 20K setting also yields the highest clean accuracy.
These results indicate that sufficient clean-data coverage is important 
for effective defense across different trigger structures. We also observe
that GARD is not sensitive to the choice of clean samples.
In Appendix Sections~G and H, we further evaluate GARD with
contaminated defense data and without clean target-class images.

\section{Conclusion}
We presented GARD, a post-hoc defense that uses adversarial perturbations to
probe and repair backdoor-induced representation vulnerabilities in
MCL models. Rather than estimating the attack target or
reconstructing an unknown trigger, GARD searches for global and local input
variations that maximally displace clean-image representations and then
realigns the perturbed representations with their clean semantic anchors.
Across six attacks and two CLIP visual backbones, GARD reduced ASR below
$1\%$ in all evaluated settings while maintaining competitive clean
accuracy. The ablation and representation analyses further showed that the
global and local objectives address complementary vulnerabilities associated
with distributed and spatially concentrated triggers. Overall, GARD demonstrates 
that identifying and suppressing backdoor-induced representation vulnerabilities
offers a practical target-free direction for defending MCL models.

{\small
\bibliographystyle{plainnat}
\bibliography{main}

@inproceedings{bansal2023cleanclip,
  author    = {Hritik Bansal and Nishad Singhi and Yu Yang and Fan Yin and Aditya Grover and Kai-Wei Chang},
  title     = {{CleanCLIP}: Mitigating Data Poisoning Attacks in Multimodal Contrastive Learning},
  booktitle = {Proceedings of the IEEE/CVF International Conference on Computer Vision},
  pages     = {112--123},
  year      = {2023}
}

@inproceedings{jia2022badencoder,
  author    = {Jinyuan Jia and Yupei Liu and Neil Zhenqiang Gong},
  title     = {{BadEncoder}: Backdoor Attacks to Pre-trained Encoders in Self-Supervised Learning},
  booktitle = {2022 IEEE Symposium on Security and Privacy},
  pages     = {2043--2059},
  year      = {2022},
  doi       = {10.1109/SP46214.2022.9833644}
}

@inproceedings{liang2024badclip,
  author    = {Siyuan Liang and Mingli Zhu and Aishan Liu and Baoyuan Wu and Xiaochun Cao and Ee-Chien Chang},
  title     = {{BadCLIP}: Dual-Embedding Guided Backdoor Attack on Multimodal Contrastive Learning},
  booktitle = {Proceedings of the IEEE/CVF Conference on Computer Vision and Pattern Recognition},
  pages     = {24645--24654},
  year      = {2024}
}

@inproceedings{liu2025badvision,
  author    = {Zhaoyi Liu and Huan Zhang},
  title     = {Stealthy Backdoor Attack in Self-Supervised Learning Vision Encoders for Large Vision Language Models},
  booktitle = {Proceedings of the IEEE/CVF Conference on Computer Vision and Pattern Recognition},
  pages     = {25060--25070},
  year      = {2025}
}

@inproceedings{yang2024safeclip,
  author    = {Wenhan Yang and Jingdong Gao and Baharan Mirzasoleiman},
  title     = {Better Safe than Sorry: Pre-training {CLIP} against Targeted Data Poisoning and Backdoor Attacks},
  booktitle = {Proceedings of the 41st International Conference on Machine Learning},
  pages     = {56096--56108},
  volume    = {235},
  series    = {Proceedings of Machine Learning Research},
  year      = {2024}
}

@inproceedings{sun2026invertune,
  author    = {Mengyuan Sun and Yu Li and Yunjie Ge and Yuchen Liu and Bo Du and Qian Wang},
  title     = {{InverTune}: A Backdoor Defense Method for Multimodal Contrastive Learning via Backdoor-Adversarial Correlation Analysis},
  booktitle = {Proceedings of the Network and Distributed System Security Symposium},
  year      = {2026}
}

@inproceedings{he2025closer,
  author    = {Shuo He and Zhifang Zhang and Feng Liu and Roy Ka-Wei Lee and Bo An and Lei Feng},
  title     = {A Closer Look at Backdoor Attacks on {CLIP}},
  booktitle = {Proceedings of the 42nd International Conference on Machine Learning},
  pages     = {22836--22852},
  volume    = {267},
  series    = {Proceedings of Machine Learning Research},
  year      = {2025}
}

@inproceedings{radford2021clip,
  author    = {Alec Radford and Jong Wook Kim and Chris Hallacy and Aditya Ramesh and Gabriel Goh and Sandhini Agarwal and Girish Sastry and Amanda Askell and Pamela Mishkin and Jack Clark and Gretchen Krueger and Ilya Sutskever},
  title     = {Learning Transferable Visual Models From Natural Language Supervision},
  booktitle = {Proceedings of the 38th International Conference on Machine Learning},
  pages     = {8748--8763},
  volume    = {139},
  series    = {Proceedings of Machine Learning Research},
  year      = {2021}
}

@inproceedings{jia2021align,
  author    = {Chao Jia and Yinfei Yang and Ye Xia and Yi-Ting Chen and Zarana Parekh and Hieu Pham and Quoc V. Le and Yun-Hsuan Sung and Zhen Li and Tom Duerig},
  title     = {Scaling Up Visual and Vision-Language Representation Learning With Noisy Text Supervision},
  booktitle = {Proceedings of the 38th International Conference on Machine Learning},
  pages     = {4904--4916},
  volume    = {139},
  series    = {Proceedings of Machine Learning Research},
  year      = {2021}
}

@inproceedings{qian2024online,
  author    = {Qi Qian and Juhua Hu},
  title     = {Online Zero-Shot Classification with {CLIP}},
  booktitle = {Computer Vision -- ECCV 2024},
  pages     = {462--477},
  volume    = {15135},
  series    = {Lecture Notes in Computer Science},
  year      = {2024},
  doi       = {10.1007/978-3-031-72980-5_27}
}

@misc{mokady2021clipcap,
  author        = {Ron Mokady and Amir Hertz and Amit H. Bermano},
  title         = {{ClipCap}: {CLIP} Prefix for Image Captioning},
  year          = {2021},
  eprint        = {2111.09734},
  archivePrefix = {arXiv},
  doi           = {10.48550/arXiv.2111.09734},
  url           = {https://arxiv.org/abs/2111.09734}
}

@inproceedings{barraco2022clipcaption,
  author    = {Manuele Barraco and Marcella Cornia and Silvia Cascianelli and Lorenzo Baraldi and Rita Cucchiara},
  title     = {The Unreasonable Effectiveness of {CLIP} Features for Image Captioning: An Experimental Analysis},
  booktitle = {Proceedings of the IEEE/CVF Conference on Computer Vision and Pattern Recognition Workshops},
  pages     = {4662--4670},
  year      = {2022}
}

@inproceedings{cho2022finegrained,
  author    = {Jaemin Cho and Seunghyun Yoon and Ajinkya Kale and Franck Dernoncourt and Trung Bui and Mohit Bansal},
  title     = {Fine-Grained Image Captioning with {CLIP} Reward},
  booktitle = {Findings of the Association for Computational Linguistics: NAACL 2022},
  pages     = {517--527},
  publisher = {Association for Computational Linguistics},
  year      = {2022},
  doi       = {10.18653/v1/2022.findings-naacl.39}
}

@inproceedings{eslami2023pubmedclip,
  author    = {Sedigheh Eslami and Christoph Meinel and Gerard de Melo},
  title     = {{PubMedCLIP}: How Much Does {CLIP} Benefit Visual Question Answering in the Medical Domain?},
  booktitle = {Findings of the Association for Computational Linguistics: EACL 2023},
  pages     = {1181--1193},
  publisher = {Association for Computational Linguistics},
  year      = {2023},
  doi       = {10.18653/v1/2023.findings-eacl.88}
}

@inproceedings{li2022blip,
  author    = {Junnan Li and Dongxu Li and Caiming Xiong and Steven Hoi},
  title     = {{BLIP}: Bootstrapping Language-Image Pre-training for Unified Vision-Language Understanding and Generation},
  booktitle = {Proceedings of the 39th International Conference on Machine Learning},
  pages     = {12888--12900},
  volume    = {162},
  series    = {Proceedings of Machine Learning Research},
  year      = {2022}
}

@inproceedings{sharma2018conceptual,
  author    = {Piyush Sharma and Nan Ding and Sebastian Goodman and Radu Soricut},
  title     = {Conceptual Captions: A Cleaned, Hypernymed, Image Alt-text Dataset For Automatic Image Captioning},
  booktitle = {Proceedings of the 56th Annual Meeting of the Association for Computational Linguistics},
  pages     = {2556--2565},
  year      = {2018},
  doi       = {10.18653/v1/P18-1238}
}

@inproceedings{deng2009imagenet,
  author    = {Jia Deng and Wei Dong and Richard Socher and Li-Jia Li and Kai Li and Li Fei-Fei},
  title     = {{ImageNet}: A Large-Scale Hierarchical Image Database},
  booktitle = {Proceedings of the IEEE Conference on Computer Vision and Pattern Recognition},
  pages     = {248--255},
  year      = {2009},
  doi       = {10.1109/CVPR.2009.5206848}
}

@article{everingham2010pascal,
  author  = {Mark Everingham and Luc Van Gool and Christopher K. I. Williams and John Winn and Andrew Zisserman},
  title   = {The Pascal Visual Object Classes ({VOC}) Challenge},
  journal = {International Journal of Computer Vision},
  volume  = {88},
  number  = {2},
  pages   = {303--338},
  year    = {2010},
  doi     = {10.1007/s11263-009-0275-4}
}

@article{gu2019badnets,
  author  = {Tianyu Gu and Brendan Dolan-Gavitt and Siddharth Garg},
  title   = {{BadNets}: Identifying Vulnerabilities in the Machine Learning Model Supply Chain},
  journal = {IEEE Access},
  volume  = {7},
  pages   = {47230--47244},
  year    = {2019},
  doi     = {10.1109/ACCESS.2019.2909068}
}

@misc{chen2017targeted,
  author        = {Xinyun Chen and Chang Liu and Bo Li and Kimberly Lu and Dawn Song},
  title         = {Targeted Backdoor Attacks on Deep Learning Systems Using Data Poisoning},
  year          = {2017},
  eprint        = {1712.05526},
  archivePrefix = {arXiv},
  doi           = {10.48550/arXiv.1712.05526},
  url           = {https://arxiv.org/abs/1712.05526}
}

@inproceedings{barni2019sig,
  author    = {Mauro Barni and Kassem Kallas and Benedetta Tondi},
  title     = {A New Backdoor Attack in {CNNs} by Training Set Corruption without Label Poisoning},
  booktitle = {2019 IEEE International Conference on Image Processing},
  pages     = {101--105},
  year      = {2019},
  doi       = {10.1109/ICIP.2019.8802997}
}

@inproceedings{singh2026par,
  author    = {Naman Deep Singh and Francesco Croce and Matthias Hein},
  title     = {Perturb and Recover: Fine-tuning for Effective Backdoor Removal from {CLIP}},
  booktitle = {Proceedings of the IEEE/CVF Conference on Computer Vision and Pattern Recognition},
  year      = {2026},
  note      = {{CVPR} 2026 Findings}
}

@inproceedings{dosovitskiy2021vit,
  author    = {Alexey Dosovitskiy and Lucas Beyer and Alexander Kolesnikov and Dirk Weissenborn and Xiaohua Zhai and Thomas Unterthiner and Mostafa Dehghani and Matthias Minderer and Georg Heigold and Sylvain Gelly and Jakob Uszkoreit and Neil Houlsby},
  title     = {An Image Is Worth 16x16 Words: Transformers for Image Recognition at Scale},
  booktitle = {International Conference on Learning Representations},
  year      = {2021}
}

@inproceedings{he2016resnet,
  author    = {Kaiming He and Xiangyu Zhang and Shaoqing Ren and Jian Sun},
  title     = {Deep Residual Learning for Image Recognition},
  booktitle = {Proceedings of the IEEE Conference on Computer Vision and Pattern Recognition},
  pages     = {770--778},
  year      = {2016}
}

@inproceedings{feng2023decree,
  author    = {Shiwei Feng and Guanhong Tao and Siyuan Cheng and Guangyu Shen and Xiangzhe Xu and Yingqi Liu and Kaiyuan Zhang and Shiqing Ma and Xiangyu Zhang},
  title     = {Detecting Backdoors in Pre-Trained Encoders},
  booktitle = {Proceedings of the IEEE/CVF Conference on Computer Vision and Pattern Recognition},
  pages     = {16352--16362},
  year      = {2023}
}

@misc{kuang2024abd,
  author        = {Junhao Kuang and Siyuan Liang and Jiawei Liang and Kuanrong Liu and Xiaochun Cao},
  title         = {Adversarial Backdoor Defense in {CLIP}},
  year          = {2024},
  eprint        = {2409.15968},
  archivePrefix = {arXiv},
  doi           = {10.48550/arXiv.2409.15968},
  url           = {https://arxiv.org/abs/2409.15968}
}

@inproceedings{madry2018towards,
  title     = {Towards Deep Learning Models Resistant to Adversarial Attacks},
  author    = {Madry, Aleksander and Makelov, Aleksandar and Schmidt, Ludwig and Tsipras, Dimitris and Vladu, Adrian},
  booktitle = {International Conference on Learning Representations},
  year      = {2018},
  url       = {https://openreview.net/forum?id=rJzIBfZAb}
}
}
\clearpage
\appendix
\setcounter{table}{0}
\setcounter{figure}{0}
\renewcommand{\thetable}{\Alph{section}\arabic{table}}
\renewcommand{\thefigure}{\Alph{section}\arabic{figure}}

\section{Diagnosing InverTune under Mismatched Trigger Structures}
\label{sec:invertune_diagnosis}

InverTune estimates the attack target using a localized adversarial patch
before performing trigger inversion and activation tuning. We examine whether
this target-identification step is sensitive to the spatial structure of the
backdoor trigger. Our analysis includes a controlled BadCLIP variant and
BadVision. Unless otherwise specified, all results in this section use a
ViT-B/32 image encoder.

\subsection{Controlled BadCLIP Variant}

\paragraph{Experimental Design.}
For this diagnostic experiment, we construct a banana-targeted BadCLIP model
and replace its original optimized local patch with a globally distributed
adversarial trigger while retaining \emph{banana} as the attack target. The
remaining attack and defense settings are kept unchanged. This comparison
isolates the effect of trigger structure on InverTune's target identification.
The modified attack achieves an ASR of 93.77\% before defense, confirming that
the substituted trigger successfully implants the intended backdoor.

\paragraph{Target-Class Identification.}
Table~\ref{tab:badclip_target_identification} reports the class most frequently
predicted after applying the adversarial patch used by InverTune. For the
original BadCLIP trigger, InverTune correctly identifies \emph{banana}, which
accounts for 76.34\% of the predictions. After the local trigger is replaced
by the global adversarial trigger, however, InverTune identifies
\emph{screen} with a prediction frequency of 72.95\%. The response of the
ground-truth target \emph{banana} becomes negligible. Thus, changing only the
trigger structure causes the target-identification stage to select an
incorrect class.

\begin{center}
\begin{minipage}{\columnwidth}
\centering
\captionof{table}{Target-class identification by InverTune for the original and
modified BadCLIP attacks. Both attacks target \emph{banana}. The frequency is
the fraction of ImageNet validation images predicted as the detected class
after applying InverTune's adversarial patch.}
\label{tab:badclip_target_identification}
\setlength{\tabcolsep}{3pt}
\begin{tabular}{lll}
\toprule
BadCLIP trigger & Detected class & Frequency (\%) \\
\midrule
Optimized local patch & banana & 76.34 \\
Global adversarial trigger & screen & 72.95 \\
\bottomrule
\end{tabular}

\end{minipage}
\end{center}

\paragraph{Effect on Backdoor Removal.}
To determine whether the identification error accounts for the subsequent
defense failure, we compare standard InverTune with an oracle variant. Standard
InverTune uses the automatically detected class \emph{screen}, whereas the
oracle variant is given the ground-truth target \emph{banana}. All other
settings are identical. As shown in
Table~\ref{tab:badclip_oracle_target}, using the incorrectly detected target
leaves an ASR of 72.62\%. Providing the correct target reduces the ASR to
3.45\%, a difference of 69.17 percentage points, while both variants retain
the same clean accuracy of 54.59\%. This controlled comparison isolates the
target label as the changed factor and shows that the target-identification
error is the primary cause of InverTune's failure on this modified attack.

\begin{center}
\begin{minipage}{\columnwidth}
\centering
\captionof{table}{Effect of the target label on InverTune for the modified BadCLIP
attack. The undefended model has an ASR of 93.77\%. The automatic setting uses
the class detected by InverTune, while the oracle setting uses the
ground-truth attack target.}
\label{tab:badclip_oracle_target}
\begin{tabular}{llcc}
\toprule
Setting & Target used & CA $\uparrow$ & ASR $\downarrow$ \\
\midrule
Automatic & screen & 54.59 & 72.62 \\
Oracle target & banana & 54.59 & 3.45 \\
\bottomrule
\end{tabular}

\end{minipage}
\end{center}

\subsection{Target Identification on BadVision}

BadVision provides a second example of a mismatch between InverTune's local
patch prior and a globally distributed trigger. BadVision targets
\emph{mushroom}, but InverTune identifies \emph{maraca}. As reported in
Table~\ref{tab:badvision_target_identification}, applying InverTune's
adversarial patch increases the prediction frequency of \emph{maraca} from
0.126\% to 8.680\%, whereas the frequency of the ground-truth target increases
only from 0.296\% to 1.012\%. The true target ranks seventh by prediction
frequency after applying the patch and ninth by absolute frequency increase.
The subsequent defense retains an ASR of 55.70\%.

\begin{center}
\begin{minipage}{\columnwidth}
\centering
\captionof{table}{Target-class responses produced by InverTune on BadVision. The
ground-truth backdoor target is \emph{mushroom}. $\Delta$ is the absolute
change in prediction frequency in percentage points after applying
InverTune's adversarial patch.}
\label{tab:badvision_target_identification}
\setlength{\tabcolsep}{3pt}
\begin{tabular}{lrrr}
\toprule
Class & Clean (\%) & Patched (\%) & $\Delta$ \\
\midrule
maraca & 0.126 & 8.680 & +8.554 \\
drake & 0.000 & 6.488 & +6.488 \\
kite & 0.044 & 2.742 & +2.698 \\
confectionery & 0.158 & 1.314 & +1.156 \\
jigsaw puzzle & 0.158 & 1.254 & +1.096 \\
\midrule
mushroom (true target) & 0.296 & 1.012 & +0.716 \\
\bottomrule
\end{tabular}

\end{minipage}
\end{center}

\paragraph{Discussion.}
Together, these results show that InverTune's effectiveness depends on the
compatibility between its target-identification prior and the trigger
structure. It successfully identifies and removes the original patch-based
BadCLIP backdoor, but its localized adversarial patch can activate unrelated
classes when the implanted trigger is globally distributed. The oracle
experiment further shows that InverTune can substantially mitigate the
modified BadCLIP backdoor once the correct target is supplied; the observed
failure therefore arises primarily from target identification rather than
from an inability to tune the compromised model under any target. These
diagnostic results motivate defenses that do not require estimating the attack
target or explicitly reconstructing its trigger.

\section{Complete Representation Visualizations}
\label{sec:complete_representation_visualizations}

\begin{figure*}[p]
    \centering
    \includegraphics[height=0.82\textheight,keepaspectratio]{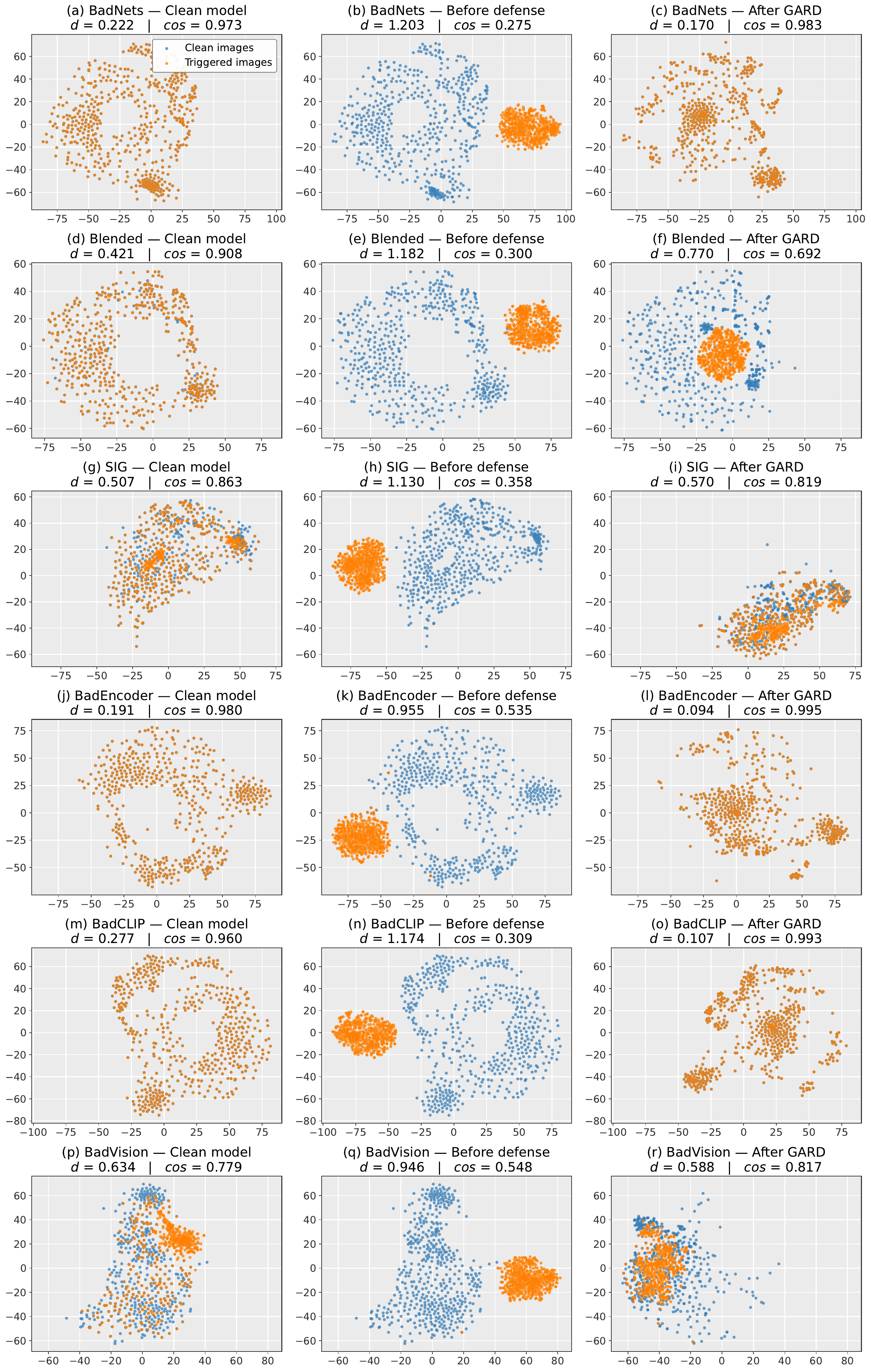}
    \caption{Complete representation visualization across six attacks on
    ViT-B/32. Each row corresponds to one attack, and the columns show the
    clean reference model, the backdoored model before defense, and the model
    after GARD. Within each attack, all representations are projected using a
    joint t-SNE embedding. The mean Euclidean distance $d$ and cosine
    similarity $\mathit{cos}$ between paired clean and triggered
    representations are computed in the original normalized feature space.}
    \label{fig:supp_complete_tsne}
\end{figure*}

We extend the representation analysis in the main paper to all six attacks on
ViT-B/32. For each attack, we compare a clean reference model, the backdoored
model before defense, and the same backdoored model after GARD.
Figure~\ref{fig:supp_complete_tsne} visualizes paired clean and triggered image
representations under these three stages.

Across all six attacks, GARD decreases the mean clean--triggered distance and
increases their mean cosine similarity relative to the undefended backdoored
model. The recovery is particularly pronounced for BadNets, BadEncoder, and
BadCLIP. For Blended and SIG, GARD substantially reduces the deviation but
does not fully recover the corresponding clean-model statistics. BadVision
shows a similar correction, with the mean distance decreasing from 0.946 to
0.588 and the mean cosine similarity increasing from 0.548 to 0.817. These
visualizations provide qualitative support for the representation
realignment effect; the reported $d$ and $\mathit{cos}$ values provide the
corresponding measurements in the original feature space.

\section{Image--Text Retrieval Results}
\label{sec:supp_retrieval}

We further evaluate the defended ViT-B/32 models on image--text retrieval.
Because ImageNet does not provide natural-language descriptions paired with
its images, GARD uses 20K clean CC3M image--text pairs for this experiment.
All other defense settings are kept unchanged. Table~\ref{tab:supp_retrieval}
reports clean retrieval accuracy (CA) and attack success rate (ASR) across six
attacks.

\begin{table*}[t]
\centering
\caption{Image--text retrieval results on ViT-B/32. CA (\%) measures clean
retrieval accuracy, and ASR (\%) measures backdoor effectiveness after
defense. GARD uses 20K clean CC3M image--text pairs because ImageNet does not
provide paired natural-language descriptions; all other settings remain
unchanged. The best and second-best defense results are shown in bold and
underlined, respectively.}
\label{tab:supp_retrieval}
\small
\setlength{\tabcolsep}{4.0pt}
\begin{tabular}{l cc cc cc cc cc cc}
\toprule
& \multicolumn{2}{c}{BadNets}
& \multicolumn{2}{c}{Blended}
& \multicolumn{2}{c}{SIG}
& \multicolumn{2}{c}{BadEncoder}
& \multicolumn{2}{c}{BadCLIP}
& \multicolumn{2}{c}{BadVision} \\
\cmidrule(lr){2-3}\cmidrule(lr){4-5}\cmidrule(lr){6-7}
\cmidrule(lr){8-9}\cmidrule(lr){10-11}\cmidrule(lr){12-13}
Method
& CA $\uparrow$ & ASR $\downarrow$
& CA $\uparrow$ & ASR $\downarrow$
& CA $\uparrow$ & ASR $\downarrow$
& CA $\uparrow$ & ASR $\downarrow$
& CA $\uparrow$ & ASR $\downarrow$
& CA $\uparrow$ & ASR $\downarrow$ \\
\midrule
No Defense
& 80.96 & 53.38 & 80.84 & 96.34 & 81.12 & 78.46
& 82.84 & 100.00 & 81.02 & 100.00 & 81.20 & 99.94 \\
FT
& 78.00 & 1.76 & 78.46 & 29.62 & 78.52 & 4.64
& 79.16 & 0.38 & 79.42 & 99.58 & 78.40 & 52.28 \\
CleanCLIP
& \underline{79.04} & 1.32 & \underline{80.36} & 0.52
& \underline{79.18} & \textbf{1.48}
& \textbf{80.16} & \underline{0.34}
& \underline{80.14} & 99.32
& \underline{79.46} & 35.16 \\
PAR
& 78.64 & \underline{0.76} & 77.32 & \underline{0.20}
& 78.36 & 6.94 & 76.98 & \textbf{0.18}
& 78.36 & 99.34 & 77.66 & \underline{7.20} \\
InverTune
& \textbf{80.86} & 5.31 & \textbf{81.00} & 3.50
& \textbf{80.84} & 7.56
& 79.08 & 0.68
& \textbf{80.40} & \underline{2.74}
& \textbf{80.24} & 50.44 \\
\textbf{GARD}
& 78.90 & \textbf{0.66} & 77.94 & \textbf{0.01}
& 73.45 & \underline{2.56}
& \underline{79.30} & 0.62
& 78.92 & \textbf{1.88}
& 76.42 & \textbf{0.02} \\
\bottomrule
\end{tabular}

\end{table*}

GARD achieves the lowest ASR on four of the six attacks, reducing ASR to
0.66\% on BadNets, 0.01\% on Blended, 1.88\% on BadCLIP, and 0.02\% on
BadVision.  Overall, GARD effectively mitigates backdoor attacks in the image--text
retrieval task. Despite using only 20K clean image--text pairs, 1/25 of the
clean data used by the other fine-tuning-based defenses, GARD maintains
competitive clean retrieval performance.

\section{Comparison with Conventional Adversarial Training}
\label{sec:gard_vs_at}

GARD uses adversarial optimization as a probe for backdoor-induced
representation sensitivity, but it differs from conventional PGD adversarial
training (PGD-AT)~\cite{madry2018towards} in both the inner maximization and
the outer objective. To adapt PGD-AT to CLIP, we replace its classification
loss with the symmetric image--text contrastive loss
$\mathcal{L}_{\mathrm{CLIP}}$. The inner loop uses random-start
$\ell_\infty$ PGD to maximize this loss over the image inputs. During outer
optimization, we minimize the CLIP loss on the resulting adversarial
image--text pairs and additionally apply the standard CLIP loss to the clean
batch to preserve image--text alignment. For a fair comparison, PGD-AT
uses the same 20K clean ImageNet image--prompt pairs and training schedule as
GARD. Both methods are trained for 10 epochs with a batch size of 64 and a
learning rate of $4.5\times10^{-6}$.

Table~\ref{tab:supp_at_comparison} compares conventional AT with GARD across
the six attacks evaluated in the main paper. Across both backbones,
conventional AT causes a pronounced and consistent reduction in clean
accuracy. Although the clean CLIP loss encourages image--text alignment,
PGD-AT still sacrifices substantial clean accuracy when optimizing
adversarial robustness. GARD instead uses the
detached clean-image representation as an explicit semantic anchor and
optimizes the perturbed views toward this reference. This clean-anchored
repair better preserves the original representation geometry, leading to
consistently higher clean accuracy than conventional AT.

The comparison also reveals that adversarial robustness does not uniformly
translate into backdoor removal. PGD-AT suppresses several attacks but remains
ineffective against others, indicating that robustness learned within a
single globally bounded perturbation space does not cover all trigger-induced
vulnerabilities. In contrast, GARD maintains low ASR across the evaluated
attacks by probing both global and local perturbation spaces and repairing
the exposed representation deviations. The results therefore distinguish
GARD from a direct application of adversarial training: its benefit arises
from clean-anchored representation repair and complementary vulnerability
probing, rather than adversarial augmentation alone.

\begin{table*}[t]
\centering
\caption{Comparison between CLIP-adapted PGD adversarial training (AT) and GARD
across six attacks and two visual backbones. CA and ASR are reported in
percent. Bold values indicate the better result between AT and GARD for each
backbone and attack.}
\label{tab:supp_at_comparison}
\small
\setlength{\tabcolsep}{3.2pt}
\begin{tabular}{ll cc cc cc cc cc cc}
\toprule
& & \multicolumn{2}{c}{BadNets}
& \multicolumn{2}{c}{Blended}
& \multicolumn{2}{c}{SIG}
& \multicolumn{2}{c}{BadEncoder}
& \multicolumn{2}{c}{BadCLIP}
& \multicolumn{2}{c}{BadVision} \\
\cmidrule(lr){3-4}\cmidrule(lr){5-6}\cmidrule(lr){7-8}
\cmidrule(lr){9-10}\cmidrule(lr){11-12}\cmidrule(lr){13-14}
Backbone & Method
& CA $\uparrow$ & ASR $\downarrow$
& CA $\uparrow$ & ASR $\downarrow$
& CA $\uparrow$ & ASR $\downarrow$
& CA $\uparrow$ & ASR $\downarrow$
& CA $\uparrow$ & ASR $\downarrow$
& CA $\uparrow$ & ASR $\downarrow$ \\
\midrule
ViT-B/32 & AT
& 49.73 & 52.87 & 57.23 & 0.02 & 48.71 & 3.24
& 50.98 & \textbf{0.02} & 50.68 & 14.80 & 52.18 & \textbf{0.03} \\
ViT-B/32 & GARD
& \textbf{59.36} & \textbf{0.10}
& \textbf{59.23} & \textbf{0.01}
& \textbf{57.61} & \textbf{0.88}
& \textbf{61.66} & 0.05
& \textbf{61.01} & \textbf{0.82}
& \textbf{60.39} & 0.63 \\
\midrule
RN50 & AT
& 46.42 & 54.03 & 47.17 & \textbf{0.01} & 48.14 & \textbf{0.20}
& 49.01 & 0.02 & 50.54 & 76.58 & 45.11 & \textbf{0.01} \\
RN50 & GARD
& \textbf{53.35} & \textbf{0.11}
& \textbf{52.79} & 0.03
& \textbf{53.44} & 0.45
& \textbf{55.76} & \textbf{0.01}
& \textbf{54.88} & \textbf{0.03}
& \textbf{53.97} & 0.05 \\
\bottomrule
\end{tabular}%

\end{table*}

\section{Detailed Experimental Settings}
\label{app:experimental_settings}

\subsection{Datasets}

For attacks trained on CC3M, the 500K image--text pairs are randomly sampled
from the full CC3M dataset~\cite{sharma2018conceptual}. The ImageNet subsets
used for defense training are randomly sampled from the ImageNet-1K training
split~\cite{deng2009imagenet}. For the 20K subset used by GARD, we maintain a
balanced class distribution by sampling 20 images from each of the 1,000
classes. We use the official ImageNet-1K validation split
(\texttt{ILSVRC2012\_img\_val}) exclusively for evaluation.

\subsection{Backdoor Attacks}
\paragraph{BadNets.}
Following the CleanCLIP protocol~\cite{bansal2023cleanclip}, BadNets is
constructed on 500K CC3M image--text pairs~\cite{sharma2018conceptual} by
poisoning 1,500 pairs (0.3\%). It uses a $16\!\times\!16$ yellow patch placed
at the bottom-right corner and is trained for 10 epochs with a batch size of
128 and a learning rate of $10^{-6}$.

\paragraph{Blended.}
Blended is constructed on the same 500K CC3M pairs by poisoning 1,500 pairs
(0.3\%) and uses a blending ratio of $0.2$. It is trained for 10 epochs with a
batch size of 128 and a learning rate of $10^{-6}$.

\paragraph{SIG.}
SIG is constructed on the same 500K CC3M pairs by poisoning 1,500 pairs
(0.3\%) and uses an amplitude of $60$ and a frequency of $6$. It is trained
for 10 epochs with a batch size of 128 and a learning rate of $10^{-6}$.

\paragraph{BadCLIP.}
BadCLIP is constructed on the same 500K CC3M pairs. We first collect natural
target captions from CC3M by selecting captions that contain \emph{mushroom}
or \emph{mushrooms}. We then randomly select 1,500 samples (0.3\%) from the
500K training set. For each selected sample, we place the optimized
$16\!\times\!16$ trigger in the bottom-right corner of the image and replace
its original caption with a target caption sampled with replacement from the
collected caption pool. The resulting poisoned samples are combined with the
remaining 498,500 clean samples to form the final training set. BadCLIP is
trained for 10 epochs with a batch size of 128 and a learning rate of
$10^{-6}$.

\paragraph{BadEncoder.}
Following its original setting, BadEncoder is constructed on
VOC~\cite{everingham2010pascal} and uses a $40\!\times\!40$ white patch placed
at the bottom-right corner.

\paragraph{BadVision.}
Following its original setting, BadVision is constructed on VOC and uses an
optimized global perturbation.

All backdoored models are initialized from the corresponding OpenAI CLIP
checkpoints~\cite{radford2021clip}.

\subsection{Backdoor Defenses}

Each defense is applied independently to the corresponding backdoored
checkpoint.

\paragraph{FT.}
FT uses 500K clean CC3M image--text pairs and is trained for 10 epochs with a
batch size of 64 and a learning rate of $4.5\times10^{-6}$.

\paragraph{CleanCLIP.}
CleanCLIP~\cite{bansal2023cleanclip} uses 500K clean CC3M image--text pairs and
is trained for 10 epochs with a batch size of 64 and a learning rate of
$4.5\times10^{-6}$.

\paragraph{PAR.}
PAR~\cite{singh2026par} uses 500K clean CC3M image--text pairs and is trained
for 10 epochs with a batch size of 64. Following
InverTune~\cite{sun2026invertune}, we set the starting and peak learning rates
to $3\times10^{-6}$ and $5\times10^{-6}$ respectively.

\paragraph{InverTune.}
InverTune~\cite{sun2026invertune} uses 50K clean ImageNet images following its
original multi-stage setting.

\paragraph{GARD.}
GARD uses a subset of 20K clean labeled ImageNet images. Each image is paired
with the prompt ``a photo of a \{class name\}'' constructed from its
ground-truth label. GARD is trained for 10 epochs with a batch size of 64 and
a learning rate of $4.5\times10^{-6}$. We set
$\lambda_{\mathrm{cm}}=1.0$ and $\lambda_g=\lambda_p=0.8$. The global
perturbation uses an $\ell_\infty$ bound of $8/255$, a step size of $1/255$,
and six optimization steps. The local perturbation uses a
fixed $32\!\times\!32$ patch centered in the image, a step size of $0.1$,
and six optimization steps.

\section{Adaptive Attack Protocols}
\label{app:adaptive_attacks}
Both implementations start from OpenAI CLIP ViT-B/32. The attacker
has access to labeled clean images and triggered inputs. 

\paragraph{Adaptive attack objective.}
At each adaptive update, the attacker combines the base attack objective
with a loss designed to resist GARD:
\begin{equation}
\mathcal{L}_{\mathrm{adaptive}}^{(k)}
= \mathcal{L}_{\mathrm{base}} + \lambda_k\mathcal{L}_{\mathrm{resist}}^{(k)},
\qquad k\in\{1,2\},
\end{equation}
where $k$ denotes the two strategies below and $\lambda_k$ controls the
strength of the added loss.

\paragraph{Simulated purification.}
In this experiment, every five outer steps the attacker copies the
current model into a surrogate and applies two GARD updates using its clean,
global-consistency, and local-consistency loss functions. After purification,
the attacker computes the targeted cross-entropy gradient on the surrogate
and adds a weighted copy of this gradient to the original encoder's base
attack gradient. This is a first-order approximation that treats the
Jacobian of the purification trajectory as the identity; it does not
backpropagate through the optimizer updates.Then we save the attacker's model 
checkpoint, apply the full GARD purification procedure, and evaluate the purified model.

\paragraph{Defense-signal evasion.}
During attack training, we use GARD's generators to
produce global and local perturbations of clean images. We incorporate
the corresponding GARD consistency losses into the base attack objective
and update the model to minimize the combined objective. By
minimizing these consistency losses during attack training, this strategy
aims to weaken the signal that guides GARD purification.

\section{Sensitivity to Contaminated Defense Data}
\label{app:contaminated_defense_data}
We examine how GARD performs when its defense set contains poisoned images.
Starting from the class-balanced 20K ImageNet subset used in the main
experiments (20 images per class), we add $n$ triggered images whose true
class differs from the attack target and assign them the target-class label.
This concentrates the contamination in the target class: at $n=1$, the
target class contains 20 clean images and one poisoned image. We evaluate
the defended ViT-B/32 model against BadCLIP as $n$ increases.

\begin{table}[t]
\centering
\caption{Effect of poisoned samples in the defense data on BadCLIP with
ViT-B/32. $n$ is the number of poisoned images added to the original 20K
clean defense set. CA and ASR are measured after GARD defense.}
\label{tab:contaminated_defense_data}
\small
\begin{tabular}{rcc}
\toprule
$n$ & ASR $\downarrow$ (\%) & CA $\uparrow$ (\%) \\
\midrule
0   & 0.82  & 61.01 \\
1   & 19.03 & 60.73 \\
2   & 68.68 & 60.14 \\
5   & 83.08 & 60.59 \\
10  & 87.28 & 60.47 \\
20  & 98.80 & 60.62 \\
100 & 99.95 & 60.79 \\
\bottomrule
\end{tabular}
\end{table}

Table~\ref{tab:contaminated_defense_data} shows that GARD still reduces ASR
with one poisoned image in the defense set, but its protection degrades
rapidly as contamination increases. ASR rises from 0.82\% without
contamination to 19.03\% with one poisoned image and 68.68\% with two,
while CA remains near 60--61\%. With 20 or more poisoned images, ASR
approaches that of the undefended model. As the number of poisoned images
grows, their triggers and incorrect target labels may increasingly reinforce
the backdoor association during defense training, eventually overwhelming
the purification effect. GARD therefore benefits from keeping the defense
set clean, but requires only a small collection of clean images rather than
the large dataset used by the fine-tuning baselines. This makes assembling
a carefully verified defense set more practical.

\section{Defense without Clean Target-Class Images}
\label{app:without_target_class}
We further test whether GARD needs clean images from the attacker's target
class. Using the same BadCLIP backdoored ViT-B/32 model as in the main
experiments, we remove all 20 target-class images from the 20K ImageNet
defense set and leave the other defense settings unchanged.

\begin{center}
\begin{minipage}{\columnwidth}
\centering
\captionof{table}{BadCLIP defense with and without clean target-class images.
Results are measured after GARD purification of the same ViT-B/32 model.}
\label{tab:without_target_class}
\small
\begin{tabular}{lcc}
\toprule
Defense data & ASR $\downarrow$ (\%) & CA $\uparrow$ (\%) \\
\midrule
Full 20K set & 0.82 & 61.01 \\
Without target class & 0.03 & 60.02 \\
\bottomrule
\end{tabular}
\end{minipage}
\end{center}

Table~\ref{tab:without_target_class} shows that GARD remains effective
without clean images from the attacker's target class. This result is
consistent with the preliminary study: global and local probes can expose
backdoor-sensitive directions using images from other classes, without
requiring clean examples of the attack target.

\section{Defense Runtime}
\label{app:defense_time}
Table~\ref{tab:defense_runtime} reports the wall-clock time of the complete
defense pipeline for each method. The timing setup uses a single NVIDIA
GeForce RTX 4090 (24\,GB). CleanCLIP, PAR, and GARD are trained for 10 epochs;
InverTune follows its multi-stage procedure, including UAP generation,
target-class identification, trigger inversion, and activation tuning with
sample selection. These measurements use the method-specific defense data
settings described above: 500K clean CC3M pairs for CleanCLIP and PAR,
50K clean ImageNet images for InverTune, and 20K for GARD.

\begin{table}[t]
\centering
\caption{Defense runtime under our experimental settings. The upper block
compares complete defense pipelines; the lower block breaks down InverTune's
runtime. All times are in minutes (lower is better). Bold indicates the
shortest total runtime.}
\label{tab:defense_runtime}
\small
\setlength{\tabcolsep}{6pt}
\renewcommand{\arraystretch}{1.12}
\begin{tabular*}{\columnwidth}{@{\extracolsep{\fill}}lr@{}}
\toprule
\textbf{Defense method} & \textbf{Total time (min)} $\downarrow$ \\
\midrule
CleanCLIP & 215.44 \\
PAR & 148.41 \\
InverTune & 93.11 \\
\textbf{GARD (ours)} & \textbf{42.50} \\
\midrule
\multicolumn{2}{@{}l}{\textit{InverTune: stage-wise breakdown}} \\
\addlinespace[2pt]
\quad UAP generation & 74.41 \\
\quad Target-class identification & 6.11 \\
\quad Trigger inversion & 11.28 \\
\quad Activation tuning (incl. selection) & 1.31 \\
\bottomrule
\end{tabular*}
\end{table}

GARD completes defense training in 42.50 minutes, compared with 215.44,
148.41, and 93.11 minutes for CleanCLIP, PAR, and InverTune, respectively.
This corresponds to speedups of $5.07\times$, $3.49\times$, and $2.19\times$
under these settings. For InverTune, UAP generation accounts for approximately
80\% of the total runtime. Despite its adversarial probing steps, GARD has
the lowest total runtime among the compared methods. 

\section{Computing Infrastructure}
\label{app:computing_infrastructure}
Experiments were run on a server equipped with eight NVIDIA GeForce RTX 4090
GPUs, each with 24\,GB of memory. The server has two Intel Xeon Platinum 8378C
CPUs at 2.80\,GHz, providing 76 physical cores and 152 logical threads in
total, and approximately 1\,TB of system memory (1007.5\,GiB recognized by the
operating system). It runs Ubuntu 22.04.5 LTS with Linux kernel
6.8.0-111-generic. The software environment consists of Python 3.11.14,
PyTorch 2.7.0+cu118, CUDA 11.8 (for both the PyTorch runtime and the system
CUDA Toolkit), and OpenCLIP (\texttt{open-clip-torch}) 3.2.0.

\end{document}